\documentclass[notitlepage,superscriptaddress,showpacs,nobalancelastpage,twocolumn,aps,prl,floatfix]{revtex4-1}
\usepackage[english]{babel}
\RequirePackage[T1]{fontenc}
\RequirePackage{times} 

\usepackage{siunitx} 
\usepackage[italicdiff]{physics}
\usepackage{amsfonts}
\usepackage{subfigure}
\usepackage{amsmath}
\usepackage{amssymb}
\usepackage{graphicx,epstopdf}
\usepackage{xspace}
\usepackage{array}
\usepackage[hidelinks]{hyperref}
\usepackage{color}
\usepackage{xcolor}
\usepackage{my_symbols}
\usepackage{CJK}
\usepackage{bm}
\usepackage{verbatim}
\usepackage{tikz}
\usepackage{float}
\usepackage{appendix}
\usetikzlibrary{calc,3d}
\usetikzlibrary{arrows}
\usetikzlibrary{snakes}
\usepackage[normalem]{ulem}

{\par}
\newcounter{mycomment}

\newcommand\rmv{\bgroup\markoverwith {\textcolor{red}{\rule[0.5ex]{2pt}{0.4pt}}}\ULon}

\begin{document}
\begin{CJK*}{UTF8}{gbsn} 
\title{A Hopfield neural network in magnetic films with natural learning}
\author{Weichao Yu (余伟超)}
\affiliation{Institute for Materials Research, Tohoku University, Sendai 980-8577, Japan}

\author{Jiang Xiao (萧江)}
\email{xiaojiang@fudan.edu.cn}
\affiliation{Department of Physics and State Key Laboratory of Surface Physics, Fudan University, Shanghai 200433, China}
\affiliation{Institute for Nanoelectronics Devices and Quantum Computing, Fudan University, Shanghai 200433, China}
\affiliation{Shanghai Research Center for Quantum Sciences, Shanghai 201315, China}

\author{Gerrit E. W. Bauer (包格瑞)}
\affiliation{WPI-AIMR, Tohoku University, Sendai 980-8577, Japan}
\affiliation{Institute for Materials Research, Tohoku University, Sendai 980-8577, Japan}
\affiliation{Zernike Institute for Advanced Materials, Groningen University, Netherlands}

\begin{abstract}

Macroscopic spin ensembles possess brain-like features such as non-linearity, plasticity, stochasticity, self-oscillations, and memory effects, and therefore offer opportunities for neuromorphic computing by spintronics devices. Here we propose a physical realization of artificial neural networks based on magnetic textures, which can update their weights intrinsically via built-in physical feedback utilizing the plasticity and large number of degrees of freedom of the magnetic domain patterns and without resource-demanding external computations. 
We demonstrate the idea by simulating the operation of a 4-node Hopfield neural network for pattern recognition.

\end{abstract}

\maketitle
\end{CJK*}

{\it Introduction.} Tremendous progress  in the last decade has propelled neuromorphic computing to the forefront of information technology. However, brain-inspired algorithms are mainly emulated by conventional  von Neumann architectures in which the computing and storage units are physically separated, thereby limiting the power of artificial intelligence algorithms. For example, the power consumption of the Alpha Go processor ($\sim \SI{1 \, }{MW}$) is 50,000 times higher than that of a human brain ($\sim \SI{20 \,}{W}$). \cite{matthewij_another_2016,goi_perspective_2020}
A more sustainable route towards artificial intelligence is an architecture with hard-wired neuromorphic functions.
Spintronic systems based on magnets share features of the brain, such as non-linearity, memory, self-oscillations, stochasticity, plasticity, high degrees of freedom, \etc. \cite{grollier_spintronic_2016}
These advantages already led to alternative computing schemes, such as the stochastic \cite{borders_integer_2019,daniels_energy-efficient_2020}, in-memory logic \cite{yu_magnetic_2020}, as well as neuromorphic \cite{grollier_neuromorphic_2020} computing.

The Artificial Neural Network (ANN) is a widely used model for neuromorphic computing, with artificial neurons and synapses that emulate biological system \cite{krose_introduction_1993}.
Neurons are devices with output signals that spike when the integrated input reaches a certain threshold. While synapses connect the neurons with tunable weights.
Both functionalities can be mimicked by spintronic devices. For example,
spin-torque oscillators can serve as artificial neurons and recognize spoken digits and vowels  \cite{torrejon_neuromorphic_2017, romera_vowel_2018}. The nonlinear dynamics of skyrmion fabrics can pre-process information for reservoir computing \cite{prychynenko_magnetic_2018, bourianoff_potential_2018,pinna_reservoir_2020}.


Memristors have a resistance that depends on its history and are widely used as artificial synapses \cite{ielmini_physics-based_2017, zidan_future_2018, rajendran_neuromorphic_2016, yu_neuro-inspired_2018}. Spintronics offers memristor functionalities by reconfigurable magnetic configurations whose resistance depends on, for example, the positions of magnetic domain walls \cite{lequeux_magnetic_2016}, the number of skyrmions \cite{huang_magnetic_2017, li_magnetic_2017, chen_compact_2018}, or the texture in antiferromagnet/ferromagnet bilayers \cite{fukami_perspective_2018}, all of which can be controlled by applied fields or currents.

\begin{figure*}[t]
\includegraphics[width=\textwidth]{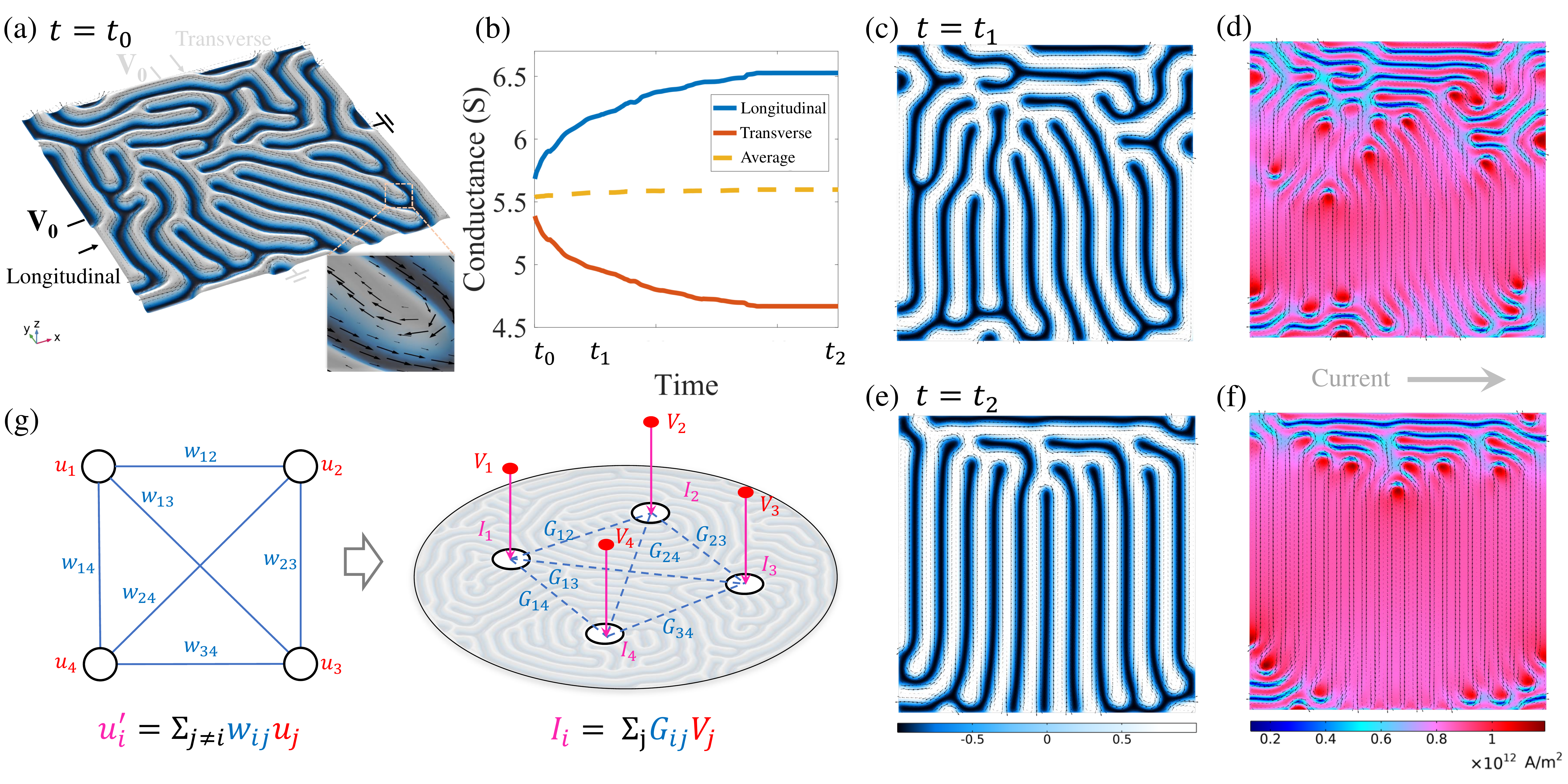}
\caption{(a) Snapshot of a maze domain structure at $t_0=0\,$ns (color code for out-of-plane magnetization and arrows for in-plane magnetization) (b) Calculated conductance $|G|$ in longitudinal ($\hbx$) and transverse ($\hby$) directions. Snapshots of magnetization (c)(e) and current density $|\mathbf{j}|$ (d)(f) distributions at $t_1=10\,$ns and $t_2=50\,$ns, respectively. An applied $V_0=0.05\,$V drives a current along the \textit{x} direction. (g) Left: Schematical Hopfield network formed by four neurons with values $u_i$ and weights $w_{ij}$. Right: Implementation of this network by a magnetic thin film with a texture that stores weights that can be trained by currents. The inputs are the voltages $V_i$ at the electrodes and the outputs are the currents $I_i$ into the film. The conductances $G_{ij}$ are equivalent to the weights in the Hopfield network.}
\label{fig1}%
\end{figure*}

In most software and hardware realizations of an ANN, the weight-updating process of the synapses is based on external algorithms, such as the back-propagation method in which external computations allocate new weights based on the results of a previous cycle. Unfortunately, this is more expensive in terms of resource and energy consumption than the inferring process itself  \cite{ceron_ai_2019}.
In this Letter we propose a platform for neuromorphic computation with superior performance that is based on the plasticity of magnetic textures. We show that an electrically conducting magnetic film may operate as a collection of artificial synapses with weights encoded by the conductances between external electrodes. The network training naturally updates the weights without requiring external computation. As a proof of principle we simulate the on-chip training and inferring (or read-out) of a 4-node Hopfield network implemented on a metallic magnetic thin film with maze spiral domains as shown in \Figure{fig1}(g).

{\it Modeling.} The dynamics of the magnetization \(\bM(\br,t)\) of a ferromagnetic film with saturation magnetization $M_s$ is governed by the Landau-Lifshitz-Gilbert (LLG) equation
\begin{equation}
\frac{\partial \mb}{\partial t}=-\gamma\mb \times \mathbf{H}_{\rm{eff}}+\alpha\mb\times\frac{\partial \mb}{\partial t}+\pmb{\tau}_{\rm{d}},
\label{LLG}
\end{equation}
where $\mb = \bM/M_s$, $\gamma$ is the gyromagnetic ratio and $\alpha$ is the Gilbert damping constant. The effective magnetic field
\begin{equation}
 \mathbf{H}_{\rm{eff}}=A\nabla^2\mb+K\mb\cdot\hat{\mathbf{z}}-D\nabla\times\mb
\label{effectivefield}
\end{equation}
consists of the exchange interaction, the perpendicular easy-axis anisotropy along $\hbz$, and a bulk-type Dzyaloshinskii-Moriya interaction (DMI) \cite{dzyaloshinsky_thermodynamic_1958,moriya_anisotropic_1960,rohart_skyrmion_2013}, parameterized by $A, K$, and $D$, respectively. The current-induced spin-transfer torque \cite{stiles_anatomy_2002,li_domain-wall_2004,seo_current-induced_2009} in \Eq{LLG}
\begin{equation}
\pmb{\tau}_{\rm{d}} = {\mu_{\rm{B}}P\ov eM_s}\bj\cdot\nabla\mb
\label{STT}
\end{equation}
is proportional to the electric current density $\bj$, where $\mu_{\rm{B}}$ is the Bohr magneton, $P$ the (conductivity) spin polarization, and $-e$ the electron charge. The dipolar interaction is not important for the energetics of  sub-micrometer scale textures and may be disregarded for the DMI-stabilized ones considered here.

The electric current density $\mathbf{j}$ is proportional to the local electric field $\bE$:
\begin{equation}
\mathbf{j}(\br)=\hat{\Sigma}[\mb(\br)]\cdot\bE(\br),
\label{ohm}
\end{equation}
where $\hat{\Sigma}[\mb]$ is the $2\times 2$ conductivity matrix of a magnetic thin film with the anisotropic magnetoresistance (AMR) \cite{thomson_xix_1857}, \ie, a local resistivity depending on the angle $\theta$ between the current flow and the local magnetization $\mb(\br)$ like
$\rho = \rho_\parallel \cos^2\theta + \rho_\perp \sin^2\theta$.  Inverting this relation, the Cartesian elements
$\Sigma_{ij}[\mb] = \sigma_\perp + \sigma_\delta m_im_j$ with $\sigma_\perp = 1/\rho_\perp$, $\sigma_\delta = 1/\rho_\parallel - 1/\rho_\perp$ and the AMR ratio $a=2\left(\rho_\parallel-\rho_\perp\right)/\left(\rho_\parallel+\rho_\perp\right)$ \cite{kruger_current-driven_2012, prychynenko_magnetic_2018, bourianoff_potential_2018, pinna_reservoir_2020}.
A large enough current-induced torque in \Eq{LLG} rotates the magnetization that in turn modulates the current distribution by \Eq{ohm}. We solve the spatiotemporal \Eqs{LLG}{ohm} self-consistently under the constraint $\nabla \cdot \mathbf{j}=0$ by the COMSOL Multiphysics \cite{comsol} finite element code.



{\it Magnetic synapse.} In magnetic films with DMI $D > 4\sqrt{AK}/\pi$ \cite{rohart_skyrmion_2013,woo_observation_2016,lan_antiferromagnetic_2017}, a spiral maze domain texture emerges \cite{viret_anisotropy_2000} as illustrated by
\Figure{fig1}(a) for a \SI{400}{nm}$\times$\SI{400}{nm} slab with thickness
$1 \, \mathrm{\mu}$m. Parameters are typical for, \eg, Pt/CoFe/MgO, but a large AMR ratio $a=150\%$ as found in Sr$_2$IrO$_4$ \cite{wang_giant_2019}, and $D=6.5\times10^{-3}\,$A. In this system many energetically nearly degenerate textures span a huge configuration space that is accessible by small variations in temperature, field or voltage. Here we focus on the overdamped regime with $\alpha=0.3$, which can be reached by  rare earth doping \cite{woltersdorf_damping_2009}. The low-energy realizations of magnetic textures are then stable in the absence of applied torque and forces, while the effects of weak pinning may be disregarded (see Supplemental Materials (SM) \cite{SM}).

Under the action of the spin-transfer torque caused by an electric voltage $V_0$ applied across the film, the maze-domain texture evolves with time, simultaneously modulating the conductance via the AMR.
\Figure{fig1}(c-f) shows snapshots of the texture with $V_0 = \SI{0.05}{V}$
applied in the $x$ direction, as in \Figure{fig1}(a).
The domains tend to align perpendicular to the current flow to minimize the spin-transfer torque, as observed in thin films of lanthanum strontium manganite  \cite{liu_current-controlled_2019}, but the sample boundaries prevent perfect alignment. The reorientation of the texture induced by the current leads to a conductance increase in the longitudinal direction ($\hbx$) and a conductance decrease in the transverse direction ($\hby$), see \Figure{fig1}(b).
The magnetic texture acts as memristors that saturates only when high voltages are applied for sufficiently long time.

{\it A self-learning Hopfield network.}
The magnetic textures equipped with electrodes establish a Hopfield network \cite{hopfield_neural_1982,rojas_neural_2013} of fully connected neurons that can recognize patterns.
\Figure{fig1}(g) sketches a Hopfield network with 4 neurons, where the synapses between different neurons are realized by the underlying magnetic film with textures. The neuron inputs are the voltages at the electrodes, which take binary values as $\qty{V_i = \pm V_0}$.
According to Kirchhoff's law $I_i = \sum_j G_{ij}[\mb]V_j$, the current through the electrodes
are governed by a (symmetric) conductance matrix $G_{ij}[\mb] = G_{ij}^0 + G'_{ij}[\mb]$ that consists of a texture-independent $\hat{G}_0$ and dependent contributions $\hat{G}'$ \footnote{Note that $\hat{G}$ is the effective conductance matrix between nodes, while $\hat{\Sigma}$ is local conductivity matrix.}.
Current conservation implies
\begin{equation}
G_{ii} = - \sum_{j\neq i} G_{ij} > 0 \qwith G_{ij} < 0 \qfor i \neq j,
\label{eqn:Gconstraint}
\end{equation}
where a current is positive when flowing into the electrodes. $\hat{G}_0$ can be measured by saturating the magnetic film by a sufficiently strong magnetic field perpendicular to the film and satisfies the same constraints as $\hat{G}$ in \Eq{eqn:Gconstraint}. The difference $\hat{G}' =\hat{G} - \hat{G}_0$ is still bound by $G'_{ii} = - \sum_{j\neq i} G'_{ij}$ but the non-diagonal elements $G'_{ij}$ can have either sign.

We may distinguish the current contributions from the texture-independent conductance $\hat{G}_0$ and that from the texture-dependent conductance $\hat{G}'$: $I_i = I_i^0 + I'_i$, where
\begin{equation}
I'_i = \sum_j G'_{ij}V_j.
\label{Kirchoff2}
\end{equation}
This completes the formulation of Hopfield network with $V_j=\pm V_0$ the neuron inputs, $G'_{ij}$ the tunable weights, and $\sign(I'_i)$ the neuron outputs.
Analogous to the the bipolar Ising spin glass model \cite{amit_spin-glass_1985}, we define the functional energy
\begin{equation}
E' = -\sum_i I'_i V_i = -\sum_{i,j}G'_{ij}V_iV_j
 = 2V_0^2\sum_{V_i\neq V_j}G'_{ij}.
\label{Energy}
\end{equation}
Physically, $-E'$ is the power consumption of the magnetic thin film (with conductance matrix $G_{ij}$) relative to that of the texture-free thin film ($G^0_{ij}$): $-E' = -E - (-E_0)$ with $-E_0 = \sum_i G_{ij}^0 V_iV_j$. With fixed $G_{ij}^0$, we know $-E_0$ in advance for arbitrary inputs $\qty{V_i}$. The  energy \Eq{Energy} therefore measures the additional Joule heating caused by the magnetic texture.

The film can be trained to memorize a pattern encoded by an array of binary values $\bV\equiv \qty{V_i}$ simply by applying the voltages $\qty{V_i}$ and let magnetization evolve. The texture adjusts itself to the current-induced spin-transfer torque such that the conductances ($\abs{G_{ij}}$ and $\abs{G'_{ij}}$) between electrode-$i$ and -$j$ increase when $V_i \neq V_j$, thereby decreasing the objective function $E'$ in \Eq{Energy}. The conductances between electrodes with the same voltage cannot be directly trained, but they tend to decrease when other conductances grow.

\begin{figure*}
\includegraphics[width=\textwidth]{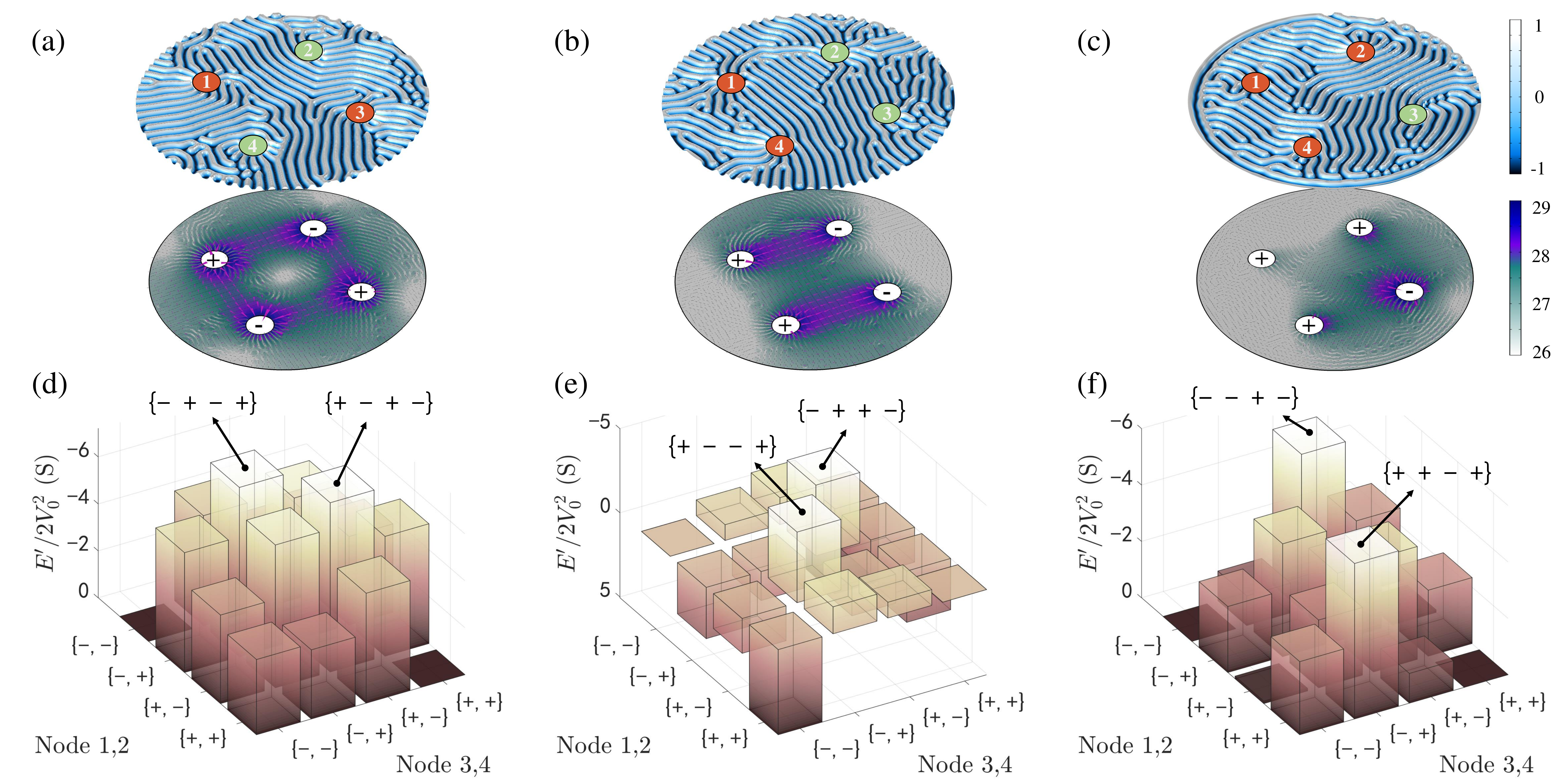}
\caption{Simulation of a 4-node Hopfield network after training for the states $\qty{V_i}= \qty{+-+\ -}$, $\qty{+--\ +}$ and $\qty{++-\ +}$, respectively. (a)-(c) Magnetization distribution (top panel, arrows for in-plane magnetization and color code for out-of-plane magnetization) and current density distribution $\log|\mathbf{j}|$ (bottom panel, magenta cones for electric current $\mathbf{j}$ on a logarithmic scale). (d)-(f) The effective energy $E'$ (\Eq{Energy}) is minimal for the trained voltage patterns.}
\label{fig:hopfield}
\end{figure*}

\Figure{fig:hopfield} shows examples of training a four-neuron network to memorize three patterns, corresponding to the voltages on the four nodes as $\{+-+\ -\}, \{+--\ +\}, \{++-\ +\}$, respectively. The top panels show the post-training texture and current density distribution.
The lower panels of \Figure{fig:hopfield} show the energies \Eq{Energy} of the trained texture when fed by all possible inputs. The energy is minimized when the trial pattern (up to an global sign change) agree with the memorized pattern. For instance, \Figure{fig:hopfield}(d) shows clear minima for the equivalent $\{+-+\ -\}$ and $\{-+-\ +\}$ states. Due to the point (rotating the pattern according to center) and mirror symmetry (flipping the voltage sign), there are 8 degenerate patterns for \Figure{fig:hopfield}(c) (2 for (a), 4 for (b)), representing 14 out of 16 possible states (that include the trivial $\{+++\ +\}$ and $\{---\ -\}$) spanned by a 4-node Hopfield network.


The memorized pattern can be retrieved by standard inferring algorithms, for example by feeding the neurons with a random initial pattern of binary voltages $\qty{V_i(0)}$ with small amplitudes that do not perturb the texture. The voltages can be then updated either asynchronously or synchronously \cite{rojas_neural_2013} with $V_i(t+1) = \sign(I'_i(t)) V_0$. The self-consistent state with $V_i = \sign(I'_i) V_0$ corresponds to the minimum of the energy function \Eq{Energy} and the memorized pattern \cite{SM}.


{\it Natural learning.} In other types of hardware-implemented synapses, such as cross-bar grid memristors \cite{kim_functional_2012,prezioso_training_2015}, the weight-updating requires learning algorithms (such as the back-propagation method) \cite{hu_associative_2015} that have to be executed externally. In contrast, the weight-updating in magnetic textures is natural and intrinsic. The positive feedback mechanism between the training current and the texture response does not require any external interference. The reinforcement of the interconnection of two ``neurons'' when activated by different voltages is analogous to Hebb's learning rule in neuroscience \cite{hartstein_self-learning_1989} stating that simultaneous activation of neurons leads to increased synaptic strength.

By dividing the continuous training process into discrete temporal slices, \ie regarding the voltages as a train of pulses, the conductance (weight) matrix evolves as
\begin{equation}
G_{ij}^{n+1} = G_{ij}^n + \Delta_{ij}^n\qty[\qty{V_i}].
\label{learningC}
\end{equation}
Here $\Delta_{ij}^n$ updates at step-$n$ the weight for the connection between node $i$ and $j$, depending mainly on the voltage difference $V_i - V_j$. The conductance matrix will not reach overall saturation
since the magnetic strips twist and move but cannot be easily created or destroyed, especially in the presence of topological defects stabilized by the DMI. The strengthening of certain connections therefore weakens others. This competition of the weight modulations is another typical feature of organic synapses that is replicated by the magnetic textures \cite{rojas_neural_2013}.

{\it Discussion.} 
The functionality and efficiency of the proposed neural network relies on the magnitude of the AMR. The AMR ratio $a$ is $\sim8\%$  in Ni$_{80}$Fe$_{20}$ thin films \cite{rijks_-plane_1997}, $-10\%/6\%$ in single-crystalline Co$_x$Fe$_{1-x}$ alloys depending on growth direction \cite{zeng_intrinsic_2020}, and $\sim80\%$ in bilayered La$_{1.2}$Sr$_{1.8}$Mn$_2$O$_7$ single crystals \cite{ning_giant_2011}.  In antiferromagnetic Sr$_2$IrO$_4$ it can reach $a \sim-160\%$ \cite{wang_giant_2019}. A larger DMI than used here ($>$\SI{0.002}{J/m^2}) reduces the pitch of the spirals and improves the efficiency of the spin-transfer torque.  The parameters and voltages are chosen optimistically to keep computation times manageable. The general idea works for less optimal materials, but at the costs of higher power dissipation and training times.
We show in the SM that the planar and anomalous Hall effects \cite{nagaosa_anomalous_2010,SM} may be disregarded because the in-plane and perpendicular magnetizations vanish on average in the absence of an external field \cite{taniguchi_spin-transfer_2015}.

The bulk-type DMI considered above generates Bloch type domain walls, that the current induced torque tends to realign such that the conductance increases.
In  materials with a dominant field-like spin-transfer torque \cite{seo_current-induced_2009} or with interfacial-type DMI \cite{fert_skyrmions_2013,rohart_skyrmion_2013}, a negative feedback reduces the conductance by applying currents (see SM \cite{SM}). Our network scheme can cooperate as well with such negative-feedback by swapping the roles of voltage and current in the inferring process and using the resistance instead of conductance as weights in the objective function.

Our proof-of-principle device can only store a 4-pixels pattern. Larger pictures can be stored by increasing the number of nodes, and multiple patterns can be memorized when the energy function has multiple local minima \cite{hopfield_neural_1982}. Scaling up the network increases the storing capacity in the form of more complex and multiple input patterns. Performance can be optimized also by the node positions or employing three dimensional textures.

{\it Conclusion.} We proposed a neural network formed by electric contacts to conducting magnets with a complex magnetization texture. Because of the plasticity of the magnetic textures, the weights of the synapses can be automatically updated during the training process via a positive feedback mechanism between the current-induced spin-transfer torque and the electrical conductance. We numerically simulate training of and retrieval from a 4-node Hopfield network on a spiral magnetic maze stabilized by DMI. The learning is natural, based on physical laws without human intervention. The concept can be generalized to other materials with ``plasticity'', such as reconfigurable ferroelectrics with conducting domain walls \cite{meier_anisotropic_2012, mcconville_ferroelectric_2020}. Our work paves the way to realize hardware-based neuromorphic computing with natural learning.

{\it Acknowledgements.}
We are grateful to Shunsuke Fukami, Hangwen Guo, Zhe Yuan and Ke Xia for fruitful discussions. This work was supported by the JSPS Kakenhi (Grant Nos. 20K14369, 19H006450). J.X. was supported by National Science Foundation of China (Grant No. 11722430) and Shanghai Municipal Science and Technology Major Project (Grant No. 2019SHZDZX01). W.Y. acknowledges the support from the State Key Laboratory of Surface Physics.


\begin{thebibliography}{55}%
\makeatletter
\providecommand \@ifxundefined [1]{%
 \@ifx{#1\undefined}
}%
\providecommand \@ifnum [1]{%
 \ifnum #1\expandafter \@firstoftwo
 \else \expandafter \@secondoftwo
 \fi
}%
\providecommand \@ifx [1]{%
 \ifx #1\expandafter \@firstoftwo
 \else \expandafter \@secondoftwo
 \fi
}%
\providecommand \natexlab [1]{#1}%
\providecommand \enquote  [1]{``#1''}%
\providecommand \bibnamefont  [1]{#1}%
\providecommand \bibfnamefont [1]{#1}%
\providecommand \citenamefont [1]{#1}%
\providecommand \href@noop [0]{\@secondoftwo}%
\providecommand \href [0]{\begingroup \@sanitize@url \@href}%
\providecommand \@href[1]{\@@startlink{#1}\@@href}%
\providecommand \@@href[1]{\endgroup#1\@@endlink}%
\providecommand \@sanitize@url [0]{\catcode `\\12\catcode `\$12\catcode
  `\&12\catcode `\#12\catcode `\^12\catcode `\_12\catcode `\%12\relax}%
\providecommand \@@startlink[1]{}%
\providecommand \@@endlink[0]{}%
\providecommand \url  [0]{\begingroup\@sanitize@url \@url }%
\providecommand \@url [1]{\endgroup\@href {#1}{\urlprefix }}%
\providecommand \urlprefix  [0]{URL }%
\providecommand \Eprint [0]{\href }%
\providecommand \doibase [0]{http://dx.doi.org/}%
\providecommand \selectlanguage [0]{\@gobble}%
\providecommand \bibinfo  [0]{\@secondoftwo}%
\providecommand \bibfield  [0]{\@secondoftwo}%
\providecommand \translation [1]{[#1]}%
\providecommand \BibitemOpen [0]{}%
\providecommand \bibitemStop [0]{}%
\providecommand \bibitemNoStop [0]{.\EOS\space}%
\providecommand \EOS [0]{\spacefactor3000\relax}%
\providecommand \BibitemShut  [1]{\csname bibitem#1\endcsname}%
\let\auto@bib@innerbib\@empty
\bibitem [{\citenamefont {{Jacques Mattheij}}()}]{matthewij_another_2016}%
  \BibitemOpen
  \bibfield  {author} {\bibinfo {author} {\bibnamefont {{Jacques Mattheij}}},\
  }\href@noop {} {\enquote {\bibinfo {title} {{Another Way Of Looking At Lee
  Sedol vs AlphaGo}},}\ }\bibinfo {howpublished}
  {\url{https://jacquesmattheij.com/another-way-of-looking-at-lee-sedol-vs-alphago/}},\
  \bibinfo {note} {online; accessed 17 March 2016}\BibitemShut {NoStop}%
\bibitem [{\citenamefont {Goi}\ \emph {et~al.}(2020)\citenamefont {Goi},
  \citenamefont {Zhang}, \citenamefont {Chen}, \citenamefont {Luan},\ and\
  \citenamefont {Gu}}]{goi_perspective_2020}%
  \BibitemOpen
  \bibfield  {author} {\bibinfo {author} {\bibfnamefont {E.}~\bibnamefont
  {Goi}}, \bibinfo {author} {\bibfnamefont {Q.}~\bibnamefont {Zhang}}, \bibinfo
  {author} {\bibfnamefont {X.}~\bibnamefont {Chen}}, \bibinfo {author}
  {\bibfnamefont {H.}~\bibnamefont {Luan}}, \ and\ \bibinfo {author}
  {\bibfnamefont {M.}~\bibnamefont {Gu}},\ }\href {\doibase
  10.1186/s43074-020-0001-6} {\bibfield  {journal} {\bibinfo  {journal}
  {PhotoniX}\ }\textbf {\bibinfo {volume} {1}},\ \bibinfo {pages} {3} (\bibinfo
  {year} {2020})}\BibitemShut {NoStop}%
\bibitem [{\citenamefont {Grollier}\ \emph {et~al.}(2016)\citenamefont
  {Grollier}, \citenamefont {Querlioz},\ and\ \citenamefont
  {Stiles}}]{grollier_spintronic_2016}%
  \BibitemOpen
  \bibfield  {author} {\bibinfo {author} {\bibfnamefont {J.}~\bibnamefont
  {Grollier}}, \bibinfo {author} {\bibfnamefont {D.}~\bibnamefont {Querlioz}},
  \ and\ \bibinfo {author} {\bibfnamefont {M.~D.}\ \bibnamefont {Stiles}},\
  }\href {\doibase 10.1109/JPROC.2016.2597152} {\bibfield  {journal} {\bibinfo
  {journal} {Proceedings of the IEEE}\ }\textbf {\bibinfo {volume} {104}},\
  \bibinfo {pages} {2024} (\bibinfo {year} {2016})}\BibitemShut {NoStop}%
\bibitem [{\citenamefont {Borders}\ \emph {et~al.}(2019)\citenamefont
  {Borders}, \citenamefont {Pervaiz}, \citenamefont {Fukami}, \citenamefont
  {Camsari}, \citenamefont {Ohno},\ and\ \citenamefont
  {Datta}}]{borders_integer_2019}%
  \BibitemOpen
  \bibfield  {author} {\bibinfo {author} {\bibfnamefont {W.~A.}\ \bibnamefont
  {Borders}}, \bibinfo {author} {\bibfnamefont {A.~Z.}\ \bibnamefont
  {Pervaiz}}, \bibinfo {author} {\bibfnamefont {S.}~\bibnamefont {Fukami}},
  \bibinfo {author} {\bibfnamefont {K.~Y.}\ \bibnamefont {Camsari}}, \bibinfo
  {author} {\bibfnamefont {H.}~\bibnamefont {Ohno}}, \ and\ \bibinfo {author}
  {\bibfnamefont {S.}~\bibnamefont {Datta}},\ }\href {\doibase
  10.1038/s41586-019-1557-9} {\bibfield  {journal} {\bibinfo  {journal}
  {Nature}\ }\textbf {\bibinfo {volume} {573}},\ \bibinfo {pages} {390}
  (\bibinfo {year} {2019})},\ \bibinfo {note} {number: 7774 Publisher: Nature
  Publishing Group}\BibitemShut {NoStop}%
\bibitem [{\citenamefont {Daniels}\ \emph {et~al.}(2020)\citenamefont
  {Daniels}, \citenamefont {Madhavan}, \citenamefont {Talatchian},
  \citenamefont {Mizrahi},\ and\ \citenamefont
  {Stiles}}]{daniels_energy-efficient_2020}%
  \BibitemOpen
  \bibfield  {author} {\bibinfo {author} {\bibfnamefont {M.~W.}\ \bibnamefont
  {Daniels}}, \bibinfo {author} {\bibfnamefont {A.}~\bibnamefont {Madhavan}},
  \bibinfo {author} {\bibfnamefont {P.}~\bibnamefont {Talatchian}}, \bibinfo
  {author} {\bibfnamefont {A.}~\bibnamefont {Mizrahi}}, \ and\ \bibinfo
  {author} {\bibfnamefont {M.~D.}\ \bibnamefont {Stiles}},\ }\href {\doibase
  10.1103/PhysRevApplied.13.034016} {\bibfield  {journal} {\bibinfo  {journal}
  {Physical Review Applied}\ }\textbf {\bibinfo {volume} {13}},\ \bibinfo
  {pages} {034016} (\bibinfo {year} {2020})},\ \bibinfo {note} {publisher:
  American Physical Society}\BibitemShut {NoStop}%
\bibitem [{\citenamefont {Yu}\ \emph {et~al.}(2020)\citenamefont {Yu},
  \citenamefont {Lan},\ and\ \citenamefont {Xiao}}]{yu_magnetic_2020}%
  \BibitemOpen
  \bibfield  {author} {\bibinfo {author} {\bibfnamefont {W.}~\bibnamefont
  {Yu}}, \bibinfo {author} {\bibfnamefont {J.}~\bibnamefont {Lan}}, \ and\
  \bibinfo {author} {\bibfnamefont {J.}~\bibnamefont {Xiao}},\ }\href {\doibase
  10.1103/PhysRevApplied.13.024055} {\bibfield  {journal} {\bibinfo  {journal}
  {Physical Review Applied}\ }\textbf {\bibinfo {volume} {13}},\ \bibinfo
  {pages} {024055} (\bibinfo {year} {2020})}\BibitemShut {NoStop}%
\bibitem [{\citenamefont {Grollier}\ \emph {et~al.}(2020)\citenamefont
  {Grollier}, \citenamefont {Querlioz}, \citenamefont {Camsari}, \citenamefont
  {Everschor-Sitte}, \citenamefont {Fukami},\ and\ \citenamefont
  {Stiles}}]{grollier_neuromorphic_2020}%
  \BibitemOpen
  \bibfield  {author} {\bibinfo {author} {\bibfnamefont {J.}~\bibnamefont
  {Grollier}}, \bibinfo {author} {\bibfnamefont {D.}~\bibnamefont {Querlioz}},
  \bibinfo {author} {\bibfnamefont {K.~Y.}\ \bibnamefont {Camsari}}, \bibinfo
  {author} {\bibfnamefont {K.}~\bibnamefont {Everschor-Sitte}}, \bibinfo
  {author} {\bibfnamefont {S.}~\bibnamefont {Fukami}}, \ and\ \bibinfo {author}
  {\bibfnamefont {M.~D.}\ \bibnamefont {Stiles}},\ }\href {\doibase
  10.1038/s41928-019-0360-9} {\bibfield  {journal} {\bibinfo  {journal} {Nature
  Electronics}\ ,\ \bibinfo {pages} {1}} (\bibinfo {year} {2020})},\ \bibinfo
  {note} {publisher: Nature Publishing Group}\BibitemShut {NoStop}%
\bibitem [{\citenamefont {Kr\"{o}se}\ and\ \citenamefont
  {Smagt}(1993)}]{krose_introduction_1993}%
  \BibitemOpen
  \bibfield  {author} {\bibinfo {author} {\bibfnamefont {B.}~\bibnamefont
  {Kr\"{o}se}}\ and\ \bibinfo {author} {\bibfnamefont {P.}~\bibnamefont
  {Smagt}},\ }\href@noop {} {\emph {\bibinfo {title} {An introduction to
  {Neural} {Networks}}}}\ (\bibinfo {year} {1993})\BibitemShut {NoStop}%
\bibitem [{\citenamefont {Torrejon}\ \emph {et~al.}(2017)\citenamefont
  {Torrejon}, \citenamefont {Riou}, \citenamefont {Araujo}, \citenamefont
  {Tsunegi}, \citenamefont {Khalsa}, \citenamefont {Querlioz}, \citenamefont
  {Bortolotti}, \citenamefont {Cros}, \citenamefont {Yakushiji}, \citenamefont
  {Fukushima}, \citenamefont {Kubota}, \citenamefont {Yuasa}, \citenamefont
  {Stiles},\ and\ \citenamefont {Grollier}}]{torrejon_neuromorphic_2017}%
  \BibitemOpen
  \bibfield  {author} {\bibinfo {author} {\bibfnamefont {J.}~\bibnamefont
  {Torrejon}}, \bibinfo {author} {\bibfnamefont {M.}~\bibnamefont {Riou}},
  \bibinfo {author} {\bibfnamefont {F.~A.}\ \bibnamefont {Araujo}}, \bibinfo
  {author} {\bibfnamefont {S.}~\bibnamefont {Tsunegi}}, \bibinfo {author}
  {\bibfnamefont {G.}~\bibnamefont {Khalsa}}, \bibinfo {author} {\bibfnamefont
  {D.}~\bibnamefont {Querlioz}}, \bibinfo {author} {\bibfnamefont
  {P.}~\bibnamefont {Bortolotti}}, \bibinfo {author} {\bibfnamefont
  {V.}~\bibnamefont {Cros}}, \bibinfo {author} {\bibfnamefont {K.}~\bibnamefont
  {Yakushiji}}, \bibinfo {author} {\bibfnamefont {A.}~\bibnamefont
  {Fukushima}}, \bibinfo {author} {\bibfnamefont {H.}~\bibnamefont {Kubota}},
  \bibinfo {author} {\bibfnamefont {S.}~\bibnamefont {Yuasa}}, \bibinfo
  {author} {\bibfnamefont {M.~D.}\ \bibnamefont {Stiles}}, \ and\ \bibinfo
  {author} {\bibfnamefont {J.}~\bibnamefont {Grollier}},\ }\href {\doibase
  10.1038/nature23011} {\bibfield  {journal} {\bibinfo  {journal} {Nature}\
  }\textbf {\bibinfo {volume} {547}},\ \bibinfo {pages} {428} (\bibinfo {year}
  {2017})}\BibitemShut {NoStop}%
\bibitem [{\citenamefont {Romera}\ \emph {et~al.}(2018)\citenamefont {Romera},
  \citenamefont {Talatchian}, \citenamefont {Tsunegi}, \citenamefont {Araujo},
  \citenamefont {Cros}, \citenamefont {Bortolotti}, \citenamefont {Trastoy},
  \citenamefont {Yakushiji}, \citenamefont {Fukushima}, \citenamefont {Kubota},
  \citenamefont {Yuasa}, \citenamefont {Ernoult}, \citenamefont
  {Vodenicarevic}, \citenamefont {Hirtzlin}, \citenamefont {Locatelli},
  \citenamefont {Querlioz},\ and\ \citenamefont
  {Grollier}}]{romera_vowel_2018}%
  \BibitemOpen
  \bibfield  {author} {\bibinfo {author} {\bibfnamefont {M.}~\bibnamefont
  {Romera}}, \bibinfo {author} {\bibfnamefont {P.}~\bibnamefont {Talatchian}},
  \bibinfo {author} {\bibfnamefont {S.}~\bibnamefont {Tsunegi}}, \bibinfo
  {author} {\bibfnamefont {F.~A.}\ \bibnamefont {Araujo}}, \bibinfo {author}
  {\bibfnamefont {V.}~\bibnamefont {Cros}}, \bibinfo {author} {\bibfnamefont
  {P.}~\bibnamefont {Bortolotti}}, \bibinfo {author} {\bibfnamefont
  {J.}~\bibnamefont {Trastoy}}, \bibinfo {author} {\bibfnamefont
  {K.}~\bibnamefont {Yakushiji}}, \bibinfo {author} {\bibfnamefont
  {A.}~\bibnamefont {Fukushima}}, \bibinfo {author} {\bibfnamefont
  {H.}~\bibnamefont {Kubota}}, \bibinfo {author} {\bibfnamefont
  {S.}~\bibnamefont {Yuasa}}, \bibinfo {author} {\bibfnamefont
  {M.}~\bibnamefont {Ernoult}}, \bibinfo {author} {\bibfnamefont
  {D.}~\bibnamefont {Vodenicarevic}}, \bibinfo {author} {\bibfnamefont
  {T.}~\bibnamefont {Hirtzlin}}, \bibinfo {author} {\bibfnamefont
  {N.}~\bibnamefont {Locatelli}}, \bibinfo {author} {\bibfnamefont
  {D.}~\bibnamefont {Querlioz}}, \ and\ \bibinfo {author} {\bibfnamefont
  {J.}~\bibnamefont {Grollier}},\ }\href {\doibase 10.1038/s41586-018-0632-y}
  {\bibfield  {journal} {\bibinfo  {journal} {Nature}\ ,\ \bibinfo {pages} {1}}
  (\bibinfo {year} {2018})}\BibitemShut {NoStop}%
\bibitem [{\citenamefont {Prychynenko}\ \emph {et~al.}(2018)\citenamefont
  {Prychynenko}, \citenamefont {Sitte}, \citenamefont {Litzius}, \citenamefont
  {Kr\"{u}ger}, \citenamefont {Bourianoff}, \citenamefont {Kl\"{a}ui},
  \citenamefont {Sinova},\ and\ \citenamefont
  {Everschor-Sitte}}]{prychynenko_magnetic_2018}%
  \BibitemOpen
  \bibfield  {author} {\bibinfo {author} {\bibfnamefont {D.}~\bibnamefont
  {Prychynenko}}, \bibinfo {author} {\bibfnamefont {M.}~\bibnamefont {Sitte}},
  \bibinfo {author} {\bibfnamefont {K.}~\bibnamefont {Litzius}}, \bibinfo
  {author} {\bibfnamefont {B.}~\bibnamefont {Kr\"{u}ger}}, \bibinfo {author}
  {\bibfnamefont {G.}~\bibnamefont {Bourianoff}}, \bibinfo {author}
  {\bibfnamefont {M.}~\bibnamefont {Kl\"{a}ui}}, \bibinfo {author}
  {\bibfnamefont {J.}~\bibnamefont {Sinova}}, \ and\ \bibinfo {author}
  {\bibfnamefont {K.}~\bibnamefont {Everschor-Sitte}},\ }\href {\doibase
  10.1103/PhysRevApplied.9.014034} {\bibfield  {journal} {\bibinfo  {journal}
  {Physical Review Applied}\ }\textbf {\bibinfo {volume} {9}},\ \bibinfo
  {pages} {014034} (\bibinfo {year} {2018})}\BibitemShut {NoStop}%
\bibitem [{\citenamefont {Bourianoff}\ \emph {et~al.}(2018)\citenamefont
  {Bourianoff}, \citenamefont {Pinna}, \citenamefont {Sitte},\ and\
  \citenamefont {Everschor-Sitte}}]{bourianoff_potential_2018}%
  \BibitemOpen
  \bibfield  {author} {\bibinfo {author} {\bibfnamefont {G.}~\bibnamefont
  {Bourianoff}}, \bibinfo {author} {\bibfnamefont {D.}~\bibnamefont {Pinna}},
  \bibinfo {author} {\bibfnamefont {M.}~\bibnamefont {Sitte}}, \ and\ \bibinfo
  {author} {\bibfnamefont {K.}~\bibnamefont {Everschor-Sitte}},\ }\href
  {\doibase 10.1063/1.5006918} {\bibfield  {journal} {\bibinfo  {journal} {AIP
  Advances}\ }\textbf {\bibinfo {volume} {8}},\ \bibinfo {pages} {055602}
  (\bibinfo {year} {2018})}\BibitemShut {NoStop}%
\bibitem [{\citenamefont {Pinna}\ \emph {et~al.}(2020)\citenamefont {Pinna},
  \citenamefont {Bourianoff},\ and\ \citenamefont
  {Everschor-Sitte}}]{pinna_reservoir_2020}%
  \BibitemOpen
  \bibfield  {author} {\bibinfo {author} {\bibfnamefont {D.}~\bibnamefont
  {Pinna}}, \bibinfo {author} {\bibfnamefont {G.}~\bibnamefont {Bourianoff}}, \
  and\ \bibinfo {author} {\bibfnamefont {K.}~\bibnamefont {Everschor-Sitte}},\
  }\href {\doibase 10.1103/PhysRevApplied.14.054020} {\bibfield  {journal}
  {\bibinfo  {journal} {Physical Review Applied}\ }\textbf {\bibinfo {volume}
  {14}},\ \bibinfo {pages} {054020} (\bibinfo {year} {2020})},\ \bibinfo {note}
  {publisher: American Physical Society}\BibitemShut {NoStop}%
\bibitem [{\citenamefont {Ielmini}\ and\ \citenamefont
  {Milo}(2017)}]{ielmini_physics-based_2017}%
  \BibitemOpen
  \bibfield  {author} {\bibinfo {author} {\bibfnamefont {D.}~\bibnamefont
  {Ielmini}}\ and\ \bibinfo {author} {\bibfnamefont {V.}~\bibnamefont {Milo}},\
  }\href {\doibase 10.1007/s10825-017-1101-9} {\bibfield  {journal} {\bibinfo
  {journal} {Journal of Computational Electronics}\ }\textbf {\bibinfo {volume}
  {16}},\ \bibinfo {pages} {1121} (\bibinfo {year} {2017})}\BibitemShut
  {NoStop}%
\bibitem [{\citenamefont {Zidan}\ \emph {et~al.}(2018)\citenamefont {Zidan},
  \citenamefont {Strachan},\ and\ \citenamefont {Lu}}]{zidan_future_2018}%
  \BibitemOpen
  \bibfield  {author} {\bibinfo {author} {\bibfnamefont {M.~A.}\ \bibnamefont
  {Zidan}}, \bibinfo {author} {\bibfnamefont {J.~P.}\ \bibnamefont {Strachan}},
  \ and\ \bibinfo {author} {\bibfnamefont {W.~D.}\ \bibnamefont {Lu}},\ }\href
  {\doibase 10.1038/s41928-017-0006-8} {\bibfield  {journal} {\bibinfo
  {journal} {Nature Electronics}\ }\textbf {\bibinfo {volume} {1}},\ \bibinfo
  {pages} {22} (\bibinfo {year} {2018})}\BibitemShut {NoStop}%
\bibitem [{\citenamefont {Rajendran}\ and\ \citenamefont
  {Alibart}(2016)}]{rajendran_neuromorphic_2016}%
  \BibitemOpen
  \bibfield  {author} {\bibinfo {author} {\bibfnamefont {B.}~\bibnamefont
  {Rajendran}}\ and\ \bibinfo {author} {\bibfnamefont {F.}~\bibnamefont
  {Alibart}},\ }\href {\doibase 10.1109/JETCAS.2016.2533298} {\bibfield
  {journal} {\bibinfo  {journal} {IEEE Journal on Emerging and Selected Topics
  in Circuits and Systems}\ }\textbf {\bibinfo {volume} {6}},\ \bibinfo {pages}
  {198} (\bibinfo {year} {2016})}\BibitemShut {NoStop}%
\bibitem [{\citenamefont {Yu}(2018)}]{yu_neuro-inspired_2018}%
  \BibitemOpen
  \bibfield  {author} {\bibinfo {author} {\bibfnamefont {S.}~\bibnamefont
  {Yu}},\ }\href {\doibase 10.1109/JPROC.2018.2790840} {\bibfield  {journal}
  {\bibinfo  {journal} {Proceedings of the IEEE}\ }\textbf {\bibinfo {volume}
  {106}},\ \bibinfo {pages} {260} (\bibinfo {year} {2018})}\BibitemShut
  {NoStop}%
\bibitem [{\citenamefont {Lequeux}\ \emph {et~al.}(2016)\citenamefont
  {Lequeux}, \citenamefont {Sampaio}, \citenamefont {Cros}, \citenamefont
  {Yakushiji}, \citenamefont {Fukushima}, \citenamefont {Matsumoto},
  \citenamefont {Kubota}, \citenamefont {Yuasa},\ and\ \citenamefont
  {Grollier}}]{lequeux_magnetic_2016}%
  \BibitemOpen
  \bibfield  {author} {\bibinfo {author} {\bibfnamefont {S.}~\bibnamefont
  {Lequeux}}, \bibinfo {author} {\bibfnamefont {J.}~\bibnamefont {Sampaio}},
  \bibinfo {author} {\bibfnamefont {V.}~\bibnamefont {Cros}}, \bibinfo {author}
  {\bibfnamefont {K.}~\bibnamefont {Yakushiji}}, \bibinfo {author}
  {\bibfnamefont {A.}~\bibnamefont {Fukushima}}, \bibinfo {author}
  {\bibfnamefont {R.}~\bibnamefont {Matsumoto}}, \bibinfo {author}
  {\bibfnamefont {H.}~\bibnamefont {Kubota}}, \bibinfo {author} {\bibfnamefont
  {S.}~\bibnamefont {Yuasa}}, \ and\ \bibinfo {author} {\bibfnamefont
  {J.}~\bibnamefont {Grollier}},\ }\href {\doibase 10.1038/srep31510}
  {\bibfield  {journal} {\bibinfo  {journal} {Scientific Reports}\ }\textbf
  {\bibinfo {volume} {6}},\ \bibinfo {pages} {31510} (\bibinfo {year}
  {2016})}\BibitemShut {NoStop}%
\bibitem [{\citenamefont {Huang}\ \emph {et~al.}(2017)\citenamefont {Huang},
  \citenamefont {Kang}, \citenamefont {Zhang}, \citenamefont {Zhou},\ and\
  \citenamefont {Zhao}}]{huang_magnetic_2017}%
  \BibitemOpen
  \bibfield  {author} {\bibinfo {author} {\bibfnamefont {Y.}~\bibnamefont
  {Huang}}, \bibinfo {author} {\bibfnamefont {W.}~\bibnamefont {Kang}},
  \bibinfo {author} {\bibfnamefont {X.}~\bibnamefont {Zhang}}, \bibinfo
  {author} {\bibfnamefont {Y.}~\bibnamefont {Zhou}}, \ and\ \bibinfo {author}
  {\bibfnamefont {W.}~\bibnamefont {Zhao}},\ }\href {\doibase
  10.1088/1361-6528/aa5838} {\bibfield  {journal} {\bibinfo  {journal}
  {Nanotechnology}\ }\textbf {\bibinfo {volume} {28}},\ \bibinfo {pages}
  {08LT02} (\bibinfo {year} {2017})}\BibitemShut {NoStop}%
\bibitem [{\citenamefont {Li}\ \emph {et~al.}(2017)\citenamefont {Li},
  \citenamefont {Kang}, \citenamefont {Huang}, \citenamefont {Zhang},
  \citenamefont {Zhou},\ and\ \citenamefont {Zhao}}]{li_magnetic_2017}%
  \BibitemOpen
  \bibfield  {author} {\bibinfo {author} {\bibfnamefont {S.}~\bibnamefont
  {Li}}, \bibinfo {author} {\bibfnamefont {W.}~\bibnamefont {Kang}}, \bibinfo
  {author} {\bibfnamefont {Y.}~\bibnamefont {Huang}}, \bibinfo {author}
  {\bibfnamefont {X.}~\bibnamefont {Zhang}}, \bibinfo {author} {\bibfnamefont
  {Y.}~\bibnamefont {Zhou}}, \ and\ \bibinfo {author} {\bibfnamefont
  {W.}~\bibnamefont {Zhao}},\ }\href {\doibase 10.1088/1361-6528/aa7af5}
  {\bibfield  {journal} {\bibinfo  {journal} {Nanotechnology}\ }\textbf
  {\bibinfo {volume} {28}},\ \bibinfo {pages} {31LT01} (\bibinfo {year}
  {2017})}\BibitemShut {NoStop}%
\bibitem [{\citenamefont {Chen}\ \emph {et~al.}(2018)\citenamefont {Chen},
  \citenamefont {Kang}, \citenamefont {Zhu}, \citenamefont {Zhang},
  \citenamefont {Lei}, \citenamefont {Zhang}, \citenamefont {Zhou},\ and\
  \citenamefont {Zhao}}]{chen_compact_2018}%
  \BibitemOpen
  \bibfield  {author} {\bibinfo {author} {\bibfnamefont {X.}~\bibnamefont
  {Chen}}, \bibinfo {author} {\bibfnamefont {W.}~\bibnamefont {Kang}}, \bibinfo
  {author} {\bibfnamefont {D.}~\bibnamefont {Zhu}}, \bibinfo {author}
  {\bibfnamefont {X.}~\bibnamefont {Zhang}}, \bibinfo {author} {\bibfnamefont
  {N.}~\bibnamefont {Lei}}, \bibinfo {author} {\bibfnamefont {Y.}~\bibnamefont
  {Zhang}}, \bibinfo {author} {\bibfnamefont {Y.}~\bibnamefont {Zhou}}, \ and\
  \bibinfo {author} {\bibfnamefont {W.}~\bibnamefont {Zhao}},\ }\href {\doibase
  10.1039/C7NR09722K} {\bibfield  {journal} {\bibinfo  {journal} {Nanoscale}\
  }\textbf {\bibinfo {volume} {10}},\ \bibinfo {pages} {6139} (\bibinfo {year}
  {2018})}\BibitemShut {NoStop}%
\bibitem [{\citenamefont {Fukami}\ and\ \citenamefont
  {Ohno}(2018)}]{fukami_perspective_2018}%
  \BibitemOpen
  \bibfield  {author} {\bibinfo {author} {\bibfnamefont {S.}~\bibnamefont
  {Fukami}}\ and\ \bibinfo {author} {\bibfnamefont {H.}~\bibnamefont {Ohno}},\
  }\href {\doibase 10.1063/1.5042317} {\bibfield  {journal} {\bibinfo
  {journal} {Journal of Applied Physics}\ }\textbf {\bibinfo {volume} {124}},\
  \bibinfo {pages} {151904} (\bibinfo {year} {2018})}\BibitemShut {NoStop}%
\bibitem [{\citenamefont {{Rodrigo Ceron}}(2019)}]{ceron_ai_2019}%
  \BibitemOpen
  \bibfield  {author} {\bibinfo {author} {\bibnamefont {{Rodrigo Ceron}}},\
  }\href
  {https://www.ibm.com/blogs/systems/ai-today-data-training-and-inferencing/}
  {{\selectlanguage {english}\enquote {\bibinfo {title} {{AI} today: {Data},
  training and inferencing},}\ }}\bibinfo {howpublished}
  {\url{https://www.ibm.com/blogs/systems/ai-today-data-training-and-inferencing/}}
  (\bibinfo {year} {2019})\BibitemShut {NoStop}%
\bibitem [{\citenamefont
  {Dzyaloshinsky}(1958)}]{dzyaloshinsky_thermodynamic_1958}%
  \BibitemOpen
  \bibfield  {author} {\bibinfo {author} {\bibfnamefont {I.}~\bibnamefont
  {Dzyaloshinsky}},\ }\href {\doibase 10.1016/0022-3697(58)90076-3} {\bibfield
  {journal} {\bibinfo  {journal} {Journal of Physics and Chemistry of Solids}\
  }\textbf {\bibinfo {volume} {4}},\ \bibinfo {pages} {241} (\bibinfo {year}
  {1958})}\BibitemShut {NoStop}%
\bibitem [{\citenamefont {Moriya}(1960)}]{moriya_anisotropic_1960}%
  \BibitemOpen
  \bibfield  {author} {\bibinfo {author} {\bibfnamefont {T.}~\bibnamefont
  {Moriya}},\ }\href {\doibase 10.1103/PhysRev.120.91} {\bibfield  {journal}
  {\bibinfo  {journal} {Physical Review}\ }\textbf {\bibinfo {volume} {120}},\
  \bibinfo {pages} {91} (\bibinfo {year} {1960})}\BibitemShut {NoStop}%
\bibitem [{\citenamefont {Rohart}\ and\ \citenamefont
  {Thiaville}(2013)}]{rohart_skyrmion_2013}%
  \BibitemOpen
  \bibfield  {author} {\bibinfo {author} {\bibfnamefont {S.}~\bibnamefont
  {Rohart}}\ and\ \bibinfo {author} {\bibfnamefont {A.}~\bibnamefont
  {Thiaville}},\ }\href {\doibase 10.1103/PhysRevB.88.184422} {\bibfield
  {journal} {\bibinfo  {journal} {Physical Review B}\ }\textbf {\bibinfo
  {volume} {88}},\ \bibinfo {pages} {184422} (\bibinfo {year}
  {2013})}\BibitemShut {NoStop}%
\bibitem [{\citenamefont {Stiles}\ and\ \citenamefont
  {Zangwill}(2002)}]{stiles_anatomy_2002}%
  \BibitemOpen
  \bibfield  {author} {\bibinfo {author} {\bibfnamefont {M.~D.}\ \bibnamefont
  {Stiles}}\ and\ \bibinfo {author} {\bibfnamefont {A.}~\bibnamefont
  {Zangwill}},\ }\href {\doibase 10.1103/PhysRevB.66.014407} {\bibfield
  {journal} {\bibinfo  {journal} {Physical Review B}\ }\textbf {\bibinfo
  {volume} {66}},\ \bibinfo {pages} {014407} (\bibinfo {year}
  {2002})}\BibitemShut {NoStop}%
\bibitem [{\citenamefont {Li}\ and\ \citenamefont
  {Zhang}(2004)}]{li_domain-wall_2004}%
  \BibitemOpen
  \bibfield  {author} {\bibinfo {author} {\bibfnamefont {Z.}~\bibnamefont
  {Li}}\ and\ \bibinfo {author} {\bibfnamefont {S.}~\bibnamefont {Zhang}},\
  }\href {\doibase 10.1103/PhysRevLett.92.207203} {\bibfield  {journal}
  {\bibinfo  {journal} {Physical Review Letters}\ }\textbf {\bibinfo {volume}
  {92}},\ \bibinfo {pages} {207203} (\bibinfo {year} {2004})}\BibitemShut
  {NoStop}%
\bibitem [{\citenamefont {Seo}\ \emph {et~al.}(2009)\citenamefont {Seo},
  \citenamefont {Lee}, \citenamefont {Yang},\ and\ \citenamefont
  {Ono}}]{seo_current-induced_2009}%
  \BibitemOpen
  \bibfield  {author} {\bibinfo {author} {\bibfnamefont {S.-M.}\ \bibnamefont
  {Seo}}, \bibinfo {author} {\bibfnamefont {K.-J.}\ \bibnamefont {Lee}},
  \bibinfo {author} {\bibfnamefont {H.}~\bibnamefont {Yang}}, \ and\ \bibinfo
  {author} {\bibfnamefont {T.}~\bibnamefont {Ono}},\ }\href {\doibase
  10.1103/PhysRevLett.102.147202} {\bibfield  {journal} {\bibinfo  {journal}
  {Physical Review Letters}\ }\textbf {\bibinfo {volume} {102}},\ \bibinfo
  {pages} {147202} (\bibinfo {year} {2009})}\BibitemShut {NoStop}%
\bibitem [{\citenamefont {Thomson}(1857)}]{thomson_xix_1857}%
  \BibitemOpen
  \bibfield  {author} {\bibinfo {author} {\bibfnamefont {W.}~\bibnamefont
  {Thomson}},\ }\href {\doibase 10.1098/rspl.1856.0144} {\bibfield  {journal}
  {\bibinfo  {journal} {Proceedings of the Royal Society of London}\ }\textbf
  {\bibinfo {volume} {8}},\ \bibinfo {pages} {546} (\bibinfo {year} {1857})},\
  \bibinfo {note} {publisher: Royal Society}\BibitemShut {NoStop}%
\bibitem [{\citenamefont {Kr\"{u}ger}(2012)}]{kruger_current-driven_2012}%
  \BibitemOpen
  \bibfield  {author} {\bibinfo {author} {\bibfnamefont {B.}~\bibnamefont
  {Kr\"{u}ger}},\ }\emph {\bibinfo {title} {Current-{Driven} {Magnetization}
  {Dynamics} : {Analytical} {Modeling} and {Numerical} {Simulation}}},\
  \href@noop {} {\bibinfo {type} {{PhD} dissertation}},\ \bibinfo  {school}
  {Universit\"{a}t Hamburg} (\bibinfo {year} {2012})\BibitemShut {NoStop}%
\bibitem [{com()}]{comsol}%
  \BibitemOpen
  \href {https://www.comsol.com/} {\enquote {\bibinfo {title} {{COMSOL}
  {Multiphysics}{\circledR} v. 5.4. www.comsol.com. {COMSOL AB}, {Stockholm},
  {Sweden}.}}\ }\BibitemShut {NoStop}%
\bibitem [{\citenamefont {Woo}\ \emph {et~al.}(2016)\citenamefont {Woo},
  \citenamefont {Litzius}, \citenamefont {Kr\"{u}ger}, \citenamefont {Im},
  \citenamefont {Caretta}, \citenamefont {Richter}, \citenamefont {Mann},
  \citenamefont {Krone}, \citenamefont {Reeve}, \citenamefont {Weigand},
  \citenamefont {Agrawal}, \citenamefont {Lemesh}, \citenamefont {Mawass},
  \citenamefont {Fischer}, \citenamefont {Kl\"{a}ui},\ and\ \citenamefont
  {Beach}}]{woo_observation_2016}%
  \BibitemOpen
  \bibfield  {author} {\bibinfo {author} {\bibfnamefont {S.}~\bibnamefont
  {Woo}}, \bibinfo {author} {\bibfnamefont {K.}~\bibnamefont {Litzius}},
  \bibinfo {author} {\bibfnamefont {B.}~\bibnamefont {Kr\"{u}ger}}, \bibinfo
  {author} {\bibfnamefont {M.-Y.}\ \bibnamefont {Im}}, \bibinfo {author}
  {\bibfnamefont {L.}~\bibnamefont {Caretta}}, \bibinfo {author} {\bibfnamefont
  {K.}~\bibnamefont {Richter}}, \bibinfo {author} {\bibfnamefont
  {M.}~\bibnamefont {Mann}}, \bibinfo {author} {\bibfnamefont {A.}~\bibnamefont
  {Krone}}, \bibinfo {author} {\bibfnamefont {R.~M.}\ \bibnamefont {Reeve}},
  \bibinfo {author} {\bibfnamefont {M.}~\bibnamefont {Weigand}}, \bibinfo
  {author} {\bibfnamefont {P.}~\bibnamefont {Agrawal}}, \bibinfo {author}
  {\bibfnamefont {I.}~\bibnamefont {Lemesh}}, \bibinfo {author} {\bibfnamefont
  {M.-A.}\ \bibnamefont {Mawass}}, \bibinfo {author} {\bibfnamefont
  {P.}~\bibnamefont {Fischer}}, \bibinfo {author} {\bibfnamefont
  {M.}~\bibnamefont {Kl\"{a}ui}}, \ and\ \bibinfo {author} {\bibfnamefont
  {G.~S.~D.}\ \bibnamefont {Beach}},\ }\href {\doibase 10.1038/nmat4593}
  {\bibfield  {journal} {\bibinfo  {journal} {Nature Materials}\ }\textbf
  {\bibinfo {volume} {15}},\ \bibinfo {pages} {501} (\bibinfo {year}
  {2016})}\BibitemShut {NoStop}%
\bibitem [{\citenamefont {Lan}\ \emph {et~al.}(2017)\citenamefont {Lan},
  \citenamefont {Yu},\ and\ \citenamefont {Xiao}}]{lan_antiferromagnetic_2017}%
  \BibitemOpen
  \bibfield  {author} {\bibinfo {author} {\bibfnamefont {J.}~\bibnamefont
  {Lan}}, \bibinfo {author} {\bibfnamefont {W.}~\bibnamefont {Yu}}, \ and\
  \bibinfo {author} {\bibfnamefont {J.}~\bibnamefont {Xiao}},\ }\href {\doibase
  10.1038/s41467-017-00265-5} {\bibfield  {journal} {\bibinfo  {journal}
  {Nature Communications}\ }\textbf {\bibinfo {volume} {8}},\ \bibinfo {pages}
  {178} (\bibinfo {year} {2017})}\BibitemShut {NoStop}%
\bibitem [{\citenamefont {Viret}\ \emph {et~al.}(2000)\citenamefont {Viret},
  \citenamefont {Samson}, \citenamefont {Warin}, \citenamefont {Marty},
  \citenamefont {Ott}, \citenamefont {S{\o}nderg{\aa}rd}, \citenamefont
  {Klein},\ and\ \citenamefont {Fermon}}]{viret_anisotropy_2000}%
  \BibitemOpen
  \bibfield  {author} {\bibinfo {author} {\bibfnamefont {M.}~\bibnamefont
  {Viret}}, \bibinfo {author} {\bibfnamefont {Y.}~\bibnamefont {Samson}},
  \bibinfo {author} {\bibfnamefont {P.}~\bibnamefont {Warin}}, \bibinfo
  {author} {\bibfnamefont {A.}~\bibnamefont {Marty}}, \bibinfo {author}
  {\bibfnamefont {F.}~\bibnamefont {Ott}}, \bibinfo {author} {\bibfnamefont
  {E.}~\bibnamefont {S{\o}nderg{\aa}rd}}, \bibinfo {author} {\bibfnamefont
  {O.}~\bibnamefont {Klein}}, \ and\ \bibinfo {author} {\bibfnamefont
  {C.}~\bibnamefont {Fermon}},\ }\href {\doibase 10.1103/PhysRevLett.85.3962}
  {\bibfield  {journal} {\bibinfo  {journal} {Physical Review Letters}\
  }\textbf {\bibinfo {volume} {85}},\ \bibinfo {pages} {3962} (\bibinfo {year}
  {2000})},\ \bibinfo {note} {publisher: American Physical Society}\BibitemShut
  {NoStop}%
\bibitem [{\citenamefont {Wang}\ \emph {et~al.}(2019)\citenamefont {Wang},
  \citenamefont {Lu}, \citenamefont {Chen}, \citenamefont {Liu}, \citenamefont
  {Yuan}, \citenamefont {Cheong}, \citenamefont {Dong},\ and\ \citenamefont
  {Liu}}]{wang_giant_2019}%
  \BibitemOpen
  \bibfield  {author} {\bibinfo {author} {\bibfnamefont {H.}~\bibnamefont
  {Wang}}, \bibinfo {author} {\bibfnamefont {C.}~\bibnamefont {Lu}}, \bibinfo
  {author} {\bibfnamefont {J.}~\bibnamefont {Chen}}, \bibinfo {author}
  {\bibfnamefont {Y.}~\bibnamefont {Liu}}, \bibinfo {author} {\bibfnamefont
  {S.~L.}\ \bibnamefont {Yuan}}, \bibinfo {author} {\bibfnamefont {S.-W.}\
  \bibnamefont {Cheong}}, \bibinfo {author} {\bibfnamefont {S.}~\bibnamefont
  {Dong}}, \ and\ \bibinfo {author} {\bibfnamefont {J.-M.}\ \bibnamefont
  {Liu}},\ }\href {\doibase 10.1038/s41467-019-10299-6} {\bibfield  {journal}
  {\bibinfo  {journal} {Nature Communications}\ }\textbf {\bibinfo {volume}
  {10}},\ \bibinfo {pages} {2280} (\bibinfo {year} {2019})},\ \bibinfo {note}
  {number: 1 Publisher: Nature Publishing Group}\BibitemShut {NoStop}%
\bibitem [{\citenamefont {Woltersdorf}\ \emph {et~al.}(2009)\citenamefont
  {Woltersdorf}, \citenamefont {Kiessling}, \citenamefont {Meyer},
  \citenamefont {Thiele},\ and\ \citenamefont
  {Back}}]{woltersdorf_damping_2009}%
  \BibitemOpen
  \bibfield  {author} {\bibinfo {author} {\bibfnamefont {G.}~\bibnamefont
  {Woltersdorf}}, \bibinfo {author} {\bibfnamefont {M.}~\bibnamefont
  {Kiessling}}, \bibinfo {author} {\bibfnamefont {G.}~\bibnamefont {Meyer}},
  \bibinfo {author} {\bibfnamefont {J.-U.}\ \bibnamefont {Thiele}}, \ and\
  \bibinfo {author} {\bibfnamefont {C.~H.}\ \bibnamefont {Back}},\ }\href
  {\doibase 10.1103/PhysRevLett.102.257602} {\bibfield  {journal} {\bibinfo
  {journal} {Physical Review Letters}\ }\textbf {\bibinfo {volume} {102}},\
  \bibinfo {pages} {257602} (\bibinfo {year} {2009})},\ \bibinfo {note}
  {publisher: American Physical Society}\BibitemShut {NoStop}%
\bibitem [{SM()}]{SM}%
  \BibitemOpen
  \href@noop {} {}\bibinfo {note} {See Supplemental Material at
  \url{hppt://link.aps.org/} for details of numerical method, material
  parameters, discussion on pinning sites, demonstration of inferring process
  and more supporting data.}\BibitemShut {Stop}%
\bibitem [{\citenamefont {Liu}\ \emph {et~al.}(2019)\citenamefont {Liu},
  \citenamefont {Wu}, \citenamefont {Zhang}, \citenamefont {Chen},
  \citenamefont {Ding}, \citenamefont {Ma}, \citenamefont {Zhang},
  \citenamefont {Sun}, \citenamefont {Tu}, \citenamefont {Wang}, \citenamefont
  {Liu}, \citenamefont {Li}, \citenamefont {Jiang}, \citenamefont {Gao},
  \citenamefont {Yu}, \citenamefont {Xiao}, \citenamefont {Duine},
  \citenamefont {Wu}, \citenamefont {Nan}, \citenamefont {Zhang},\ and\
  \citenamefont {Yu}}]{liu_current-controlled_2019}%
  \BibitemOpen
  \bibfield  {author} {\bibinfo {author} {\bibfnamefont {C.}~\bibnamefont
  {Liu}}, \bibinfo {author} {\bibfnamefont {S.}~\bibnamefont {Wu}}, \bibinfo
  {author} {\bibfnamefont {J.}~\bibnamefont {Zhang}}, \bibinfo {author}
  {\bibfnamefont {J.}~\bibnamefont {Chen}}, \bibinfo {author} {\bibfnamefont
  {J.}~\bibnamefont {Ding}}, \bibinfo {author} {\bibfnamefont {J.}~\bibnamefont
  {Ma}}, \bibinfo {author} {\bibfnamefont {Y.}~\bibnamefont {Zhang}}, \bibinfo
  {author} {\bibfnamefont {Y.}~\bibnamefont {Sun}}, \bibinfo {author}
  {\bibfnamefont {S.}~\bibnamefont {Tu}}, \bibinfo {author} {\bibfnamefont
  {H.}~\bibnamefont {Wang}}, \bibinfo {author} {\bibfnamefont {P.}~\bibnamefont
  {Liu}}, \bibinfo {author} {\bibfnamefont {C.}~\bibnamefont {Li}}, \bibinfo
  {author} {\bibfnamefont {Y.}~\bibnamefont {Jiang}}, \bibinfo {author}
  {\bibfnamefont {P.}~\bibnamefont {Gao}}, \bibinfo {author} {\bibfnamefont
  {D.}~\bibnamefont {Yu}}, \bibinfo {author} {\bibfnamefont {J.}~\bibnamefont
  {Xiao}}, \bibinfo {author} {\bibfnamefont {R.}~\bibnamefont {Duine}},
  \bibinfo {author} {\bibfnamefont {M.}~\bibnamefont {Wu}}, \bibinfo {author}
  {\bibfnamefont {C.-W.}\ \bibnamefont {Nan}}, \bibinfo {author} {\bibfnamefont
  {J.}~\bibnamefont {Zhang}}, \ and\ \bibinfo {author} {\bibfnamefont
  {H.}~\bibnamefont {Yu}},\ }\href {\doibase 10.1038/s41565-019-0429-7}
  {\bibfield  {journal} {\bibinfo  {journal} {Nature Nanotechnology}\ }\textbf
  {\bibinfo {volume} {14}},\ \bibinfo {pages} {691} (\bibinfo {year} {2019})},\
  \bibinfo {note} {number: 7 Publisher: Nature Publishing Group}\BibitemShut
  {NoStop}%
\bibitem [{\citenamefont {Hopfield}(1982)}]{hopfield_neural_1982}%
  \BibitemOpen
  \bibfield  {author} {\bibinfo {author} {\bibfnamefont {J.~J.}\ \bibnamefont
  {Hopfield}},\ }\href {\doibase 10.1073/pnas.79.8.2554} {\bibfield  {journal}
  {\bibinfo  {journal} {Proceedings of the National Academy of Sciences}\
  }\textbf {\bibinfo {volume} {79}},\ \bibinfo {pages} {2554} (\bibinfo {year}
  {1982})}\BibitemShut {NoStop}%
\bibitem [{\citenamefont {Rojas}(2013)}]{rojas_neural_2013}%
  \BibitemOpen
  \bibfield  {author} {\bibinfo {author} {\bibfnamefont {R.}~\bibnamefont
  {Rojas}},\ }\href@noop {} {{\selectlanguage {english}\emph {\bibinfo {title}
  {Neural {Networks}: {A} {Systematic} {Introduction}}}}}\ (\bibinfo
  {publisher} {Springer Science \& Business Media},\ \bibinfo {year} {2013})\
  \bibinfo {note} {google-Books-ID: 4rESBwAAQBAJ}\BibitemShut {NoStop}%
\bibitem [{Note1()}]{Note1}%
  \BibitemOpen
  \bibinfo {note} {Note that $\protect \mathaccentV {hat}05E{G}$ is the
  effective conductance matrix between nodes, while $\protect \mathaccentV
  {hat}05E{\Sigma }$ is local conductivity matrix.}\BibitemShut {Stop}%
\bibitem [{\citenamefont {Amit}\ \emph {et~al.}(1985)\citenamefont {Amit},
  \citenamefont {Gutfreund},\ and\ \citenamefont
  {Sompolinsky}}]{amit_spin-glass_1985}%
  \BibitemOpen
  \bibfield  {author} {\bibinfo {author} {\bibfnamefont {D.~J.}\ \bibnamefont
  {Amit}}, \bibinfo {author} {\bibfnamefont {H.}~\bibnamefont {Gutfreund}}, \
  and\ \bibinfo {author} {\bibfnamefont {H.}~\bibnamefont {Sompolinsky}},\
  }\href {\doibase 10.1103/PhysRevA.32.1007} {\bibfield  {journal} {\bibinfo
  {journal} {Physical Review A}\ }\textbf {\bibinfo {volume} {32}},\ \bibinfo
  {pages} {1007} (\bibinfo {year} {1985})},\ \bibinfo {note} {publisher:
  American Physical Society}\BibitemShut {NoStop}%
\bibitem [{\citenamefont {Kim}\ \emph {et~al.}(2012)\citenamefont {Kim},
  \citenamefont {Gaba}, \citenamefont {Wheeler}, \citenamefont {Cruz-Albrecht},
  \citenamefont {Hussain}, \citenamefont {Srinivasa},\ and\ \citenamefont
  {Lu}}]{kim_functional_2012}%
  \BibitemOpen
  \bibfield  {author} {\bibinfo {author} {\bibfnamefont {K.-H.}\ \bibnamefont
  {Kim}}, \bibinfo {author} {\bibfnamefont {S.}~\bibnamefont {Gaba}}, \bibinfo
  {author} {\bibfnamefont {D.}~\bibnamefont {Wheeler}}, \bibinfo {author}
  {\bibfnamefont {J.~M.}\ \bibnamefont {Cruz-Albrecht}}, \bibinfo {author}
  {\bibfnamefont {T.}~\bibnamefont {Hussain}}, \bibinfo {author} {\bibfnamefont
  {N.}~\bibnamefont {Srinivasa}}, \ and\ \bibinfo {author} {\bibfnamefont
  {W.}~\bibnamefont {Lu}},\ }\href {\doibase 10.1021/nl203687n} {\bibfield
  {journal} {\bibinfo  {journal} {Nano Letters}\ }\textbf {\bibinfo {volume}
  {12}},\ \bibinfo {pages} {389} (\bibinfo {year} {2012})},\ \bibinfo {note}
  {publisher: American Chemical Society}\BibitemShut {NoStop}%
\bibitem [{\citenamefont {Prezioso}\ \emph {et~al.}(2015)\citenamefont
  {Prezioso}, \citenamefont {Merrikh-Bayat}, \citenamefont {Hoskins},
  \citenamefont {Adam}, \citenamefont {Likharev},\ and\ \citenamefont
  {Strukov}}]{prezioso_training_2015}%
  \BibitemOpen
  \bibfield  {author} {\bibinfo {author} {\bibfnamefont {M.}~\bibnamefont
  {Prezioso}}, \bibinfo {author} {\bibfnamefont {F.}~\bibnamefont
  {Merrikh-Bayat}}, \bibinfo {author} {\bibfnamefont {B.~D.}\ \bibnamefont
  {Hoskins}}, \bibinfo {author} {\bibfnamefont {G.~C.}\ \bibnamefont {Adam}},
  \bibinfo {author} {\bibfnamefont {K.~K.}\ \bibnamefont {Likharev}}, \ and\
  \bibinfo {author} {\bibfnamefont {D.~B.}\ \bibnamefont {Strukov}},\ }\href
  {\doibase 10.1038/nature14441} {\bibfield  {journal} {\bibinfo  {journal}
  {Nature}\ }\textbf {\bibinfo {volume} {521}},\ \bibinfo {pages} {61}
  (\bibinfo {year} {2015})}\BibitemShut {NoStop}%
\bibitem [{\citenamefont {Hu}\ \emph {et~al.}(2015)\citenamefont {Hu},
  \citenamefont {Liu}, \citenamefont {Liu}, \citenamefont {Chen}, \citenamefont
  {Wang}, \citenamefont {Yu}, \citenamefont {Deng}, \citenamefont {Yin},\ and\
  \citenamefont {Hosaka}}]{hu_associative_2015}%
  \BibitemOpen
  \bibfield  {author} {\bibinfo {author} {\bibfnamefont {S.~G.}\ \bibnamefont
  {Hu}}, \bibinfo {author} {\bibfnamefont {Y.}~\bibnamefont {Liu}}, \bibinfo
  {author} {\bibfnamefont {Z.}~\bibnamefont {Liu}}, \bibinfo {author}
  {\bibfnamefont {T.~P.}\ \bibnamefont {Chen}}, \bibinfo {author}
  {\bibfnamefont {J.~J.}\ \bibnamefont {Wang}}, \bibinfo {author}
  {\bibfnamefont {Q.}~\bibnamefont {Yu}}, \bibinfo {author} {\bibfnamefont
  {L.~J.}\ \bibnamefont {Deng}}, \bibinfo {author} {\bibfnamefont
  {Y.}~\bibnamefont {Yin}}, \ and\ \bibinfo {author} {\bibfnamefont
  {S.}~\bibnamefont {Hosaka}},\ }\href {\doibase 10.1038/ncomms8522} {\bibfield
   {journal} {\bibinfo  {journal} {Nature Communications}\ }\textbf {\bibinfo
  {volume} {6}},\ \bibinfo {pages} {7522} (\bibinfo {year} {2015})},\ \bibinfo
  {note} {number: 1 Publisher: Nature Publishing Group}\BibitemShut {NoStop}%
\bibitem [{\citenamefont {Hartstein}\ and\ \citenamefont
  {Koch}(1989)}]{hartstein_self-learning_1989}%
  \BibitemOpen
  \bibfield  {author} {\bibinfo {author} {\bibfnamefont {A.}~\bibnamefont
  {Hartstein}}\ and\ \bibinfo {author} {\bibfnamefont {R.~H.}\ \bibnamefont
  {Koch}},\ }in\ \href
  {http://papers.nips.cc/paper/189-a-self-learning-neural-network.pdf} {\emph
  {\bibinfo {booktitle} {Advances in {Neural} {Information} {Processing}
  {Systems} 1}}},\ \bibinfo {editor} {edited by\ \bibinfo {editor}
  {\bibfnamefont {D.~S.}\ \bibnamefont {Touretzky}}}\ (\bibinfo  {publisher}
  {Morgan-Kaufmann},\ \bibinfo {year} {1989})\ pp.\ \bibinfo {pages}
  {769--776}\BibitemShut {NoStop}%
\bibitem [{\citenamefont {Rijks}\ \emph {et~al.}(1997)\citenamefont {Rijks},
  \citenamefont {Lenczowski}, \citenamefont {Coehoorn},\ and\ \citenamefont
  {de~Jonge}}]{rijks_-plane_1997}%
  \BibitemOpen
  \bibfield  {author} {\bibinfo {author} {\bibfnamefont {T.~G. S.~M.}\
  \bibnamefont {Rijks}}, \bibinfo {author} {\bibfnamefont {S.~K.~J.}\
  \bibnamefont {Lenczowski}}, \bibinfo {author} {\bibfnamefont
  {R.}~\bibnamefont {Coehoorn}}, \ and\ \bibinfo {author} {\bibfnamefont
  {W.~J.~M.}\ \bibnamefont {de~Jonge}},\ }\href {\doibase
  10.1103/PhysRevB.56.362} {\bibfield  {journal} {\bibinfo  {journal} {Physical
  Review B}\ }\textbf {\bibinfo {volume} {56}},\ \bibinfo {pages} {362}
  (\bibinfo {year} {1997})},\ \bibinfo {note} {publisher: American Physical
  Society}\BibitemShut {NoStop}%
\bibitem [{\citenamefont {Zeng}\ \emph {et~al.}(2020)\citenamefont {Zeng},
  \citenamefont {Ren}, \citenamefont {Li}, \citenamefont {Zeng}, \citenamefont
  {Jia}, \citenamefont {Miao}, \citenamefont {Hoffmann}, \citenamefont {Zhang},
  \citenamefont {Wu},\ and\ \citenamefont {Yuan}}]{zeng_intrinsic_2020}%
  \BibitemOpen
  \bibfield  {author} {\bibinfo {author} {\bibfnamefont {F.}~\bibnamefont
  {Zeng}}, \bibinfo {author} {\bibfnamefont {Z.}~\bibnamefont {Ren}}, \bibinfo
  {author} {\bibfnamefont {Y.}~\bibnamefont {Li}}, \bibinfo {author}
  {\bibfnamefont {J.}~\bibnamefont {Zeng}}, \bibinfo {author} {\bibfnamefont
  {M.}~\bibnamefont {Jia}}, \bibinfo {author} {\bibfnamefont {J.}~\bibnamefont
  {Miao}}, \bibinfo {author} {\bibfnamefont {A.}~\bibnamefont {Hoffmann}},
  \bibinfo {author} {\bibfnamefont {W.}~\bibnamefont {Zhang}}, \bibinfo
  {author} {\bibfnamefont {Y.}~\bibnamefont {Wu}}, \ and\ \bibinfo {author}
  {\bibfnamefont {Z.}~\bibnamefont {Yuan}},\ }\href {\doibase
  10.1103/PhysRevLett.125.097201} {\bibfield  {journal} {\bibinfo  {journal}
  {Physical Review Letters}\ }\textbf {\bibinfo {volume} {125}},\ \bibinfo
  {pages} {097201} (\bibinfo {year} {2020})},\ \bibinfo {note} {publisher:
  American Physical Society}\BibitemShut {NoStop}%
\bibitem [{\citenamefont {Ning}\ \emph {et~al.}(2011)\citenamefont {Ning},
  \citenamefont {Qu}, \citenamefont {Zou}, \citenamefont {Ling}, \citenamefont
  {Zhang}, \citenamefont {Xi}, \citenamefont {Du}, \citenamefont {Li},\ and\
  \citenamefont {Zhang}}]{ning_giant_2011}%
  \BibitemOpen
  \bibfield  {author} {\bibinfo {author} {\bibfnamefont {W.}~\bibnamefont
  {Ning}}, \bibinfo {author} {\bibfnamefont {Z.}~\bibnamefont {Qu}}, \bibinfo
  {author} {\bibfnamefont {Y.-M.}\ \bibnamefont {Zou}}, \bibinfo {author}
  {\bibfnamefont {L.-S.}\ \bibnamefont {Ling}}, \bibinfo {author}
  {\bibfnamefont {L.}~\bibnamefont {Zhang}}, \bibinfo {author} {\bibfnamefont
  {C.-Y.}\ \bibnamefont {Xi}}, \bibinfo {author} {\bibfnamefont {H.-F.}\
  \bibnamefont {Du}}, \bibinfo {author} {\bibfnamefont {R.-W.}\ \bibnamefont
  {Li}}, \ and\ \bibinfo {author} {\bibfnamefont {Y.-H.}\ \bibnamefont
  {Zhang}},\ }\href {\doibase 10.1063/1.3593486} {\bibfield  {journal}
  {\bibinfo  {journal} {Applied Physics Letters}\ }\textbf {\bibinfo {volume}
  {98}},\ \bibinfo {pages} {212503} (\bibinfo {year} {2011})},\ \bibinfo {note}
  {publisher: American Institute of Physics}\BibitemShut {NoStop}%
\bibitem [{\citenamefont {Nagaosa}\ \emph {et~al.}(2010)\citenamefont
  {Nagaosa}, \citenamefont {Sinova}, \citenamefont {Onoda}, \citenamefont
  {MacDonald},\ and\ \citenamefont {Ong}}]{nagaosa_anomalous_2010}%
  \BibitemOpen
  \bibfield  {author} {\bibinfo {author} {\bibfnamefont {N.}~\bibnamefont
  {Nagaosa}}, \bibinfo {author} {\bibfnamefont {J.}~\bibnamefont {Sinova}},
  \bibinfo {author} {\bibfnamefont {S.}~\bibnamefont {Onoda}}, \bibinfo
  {author} {\bibfnamefont {A.~H.}\ \bibnamefont {MacDonald}}, \ and\ \bibinfo
  {author} {\bibfnamefont {N.~P.}\ \bibnamefont {Ong}},\ }\href {\doibase
  10.1103/RevModPhys.82.1539} {\bibfield  {journal} {\bibinfo  {journal}
  {Reviews of Modern Physics}\ }\textbf {\bibinfo {volume} {82}},\ \bibinfo
  {pages} {1539} (\bibinfo {year} {2010})},\ \bibinfo {note} {publisher:
  American Physical Society}\BibitemShut {NoStop}%
\bibitem [{\citenamefont {Taniguchi}\ \emph {et~al.}(2015)\citenamefont
  {Taniguchi}, \citenamefont {Grollier},\ and\ \citenamefont
  {Stiles}}]{taniguchi_spin-transfer_2015}%
  \BibitemOpen
  \bibfield  {author} {\bibinfo {author} {\bibfnamefont {T.}~\bibnamefont
  {Taniguchi}}, \bibinfo {author} {\bibfnamefont {J.}~\bibnamefont {Grollier}},
  \ and\ \bibinfo {author} {\bibfnamefont {M.}~\bibnamefont {Stiles}},\ }\href
  {\doibase 10.1103/PhysRevApplied.3.044001} {\bibfield  {journal} {\bibinfo
  {journal} {Physical Review Applied}\ }\textbf {\bibinfo {volume} {3}},\
  \bibinfo {pages} {044001} (\bibinfo {year} {2015})},\ \bibinfo {note}
  {publisher: American Physical Society}\BibitemShut {NoStop}%
\bibitem [{\citenamefont {Fert}\ \emph {et~al.}(2013)\citenamefont {Fert},
  \citenamefont {Cros},\ and\ \citenamefont {Sampaio}}]{fert_skyrmions_2013}%
  \BibitemOpen
  \bibfield  {author} {\bibinfo {author} {\bibfnamefont {A.}~\bibnamefont
  {Fert}}, \bibinfo {author} {\bibfnamefont {V.}~\bibnamefont {Cros}}, \ and\
  \bibinfo {author} {\bibfnamefont {J.}~\bibnamefont {Sampaio}},\ }\href
  {\doibase 10.1038/nnano.2013.29} {{\selectlanguage {english}\enquote
  {\bibinfo {title} {Skyrmions on the track},}\ }} (\bibinfo {year}
  {2013})\BibitemShut {NoStop}%
\bibitem [{\citenamefont {Meier}\ \emph {et~al.}(2012)\citenamefont {Meier},
  \citenamefont {Seidel}, \citenamefont {Cano}, \citenamefont {Delaney},
  \citenamefont {Kumagai}, \citenamefont {Mostovoy}, \citenamefont {Spaldin},
  \citenamefont {Ramesh},\ and\ \citenamefont
  {Fiebig}}]{meier_anisotropic_2012}%
  \BibitemOpen
  \bibfield  {author} {\bibinfo {author} {\bibfnamefont {D.}~\bibnamefont
  {Meier}}, \bibinfo {author} {\bibfnamefont {J.}~\bibnamefont {Seidel}},
  \bibinfo {author} {\bibfnamefont {A.}~\bibnamefont {Cano}}, \bibinfo {author}
  {\bibfnamefont {K.}~\bibnamefont {Delaney}}, \bibinfo {author} {\bibfnamefont
  {Y.}~\bibnamefont {Kumagai}}, \bibinfo {author} {\bibfnamefont
  {M.}~\bibnamefont {Mostovoy}}, \bibinfo {author} {\bibfnamefont {N.~A.}\
  \bibnamefont {Spaldin}}, \bibinfo {author} {\bibfnamefont {R.}~\bibnamefont
  {Ramesh}}, \ and\ \bibinfo {author} {\bibfnamefont {M.}~\bibnamefont
  {Fiebig}},\ }\href {\doibase 10.1038/nmat3249} {\bibfield  {journal}
  {\bibinfo  {journal} {Nature Materials}\ }\textbf {\bibinfo {volume} {11}},\
  \bibinfo {pages} {284} (\bibinfo {year} {2012})},\ \bibinfo {note} {number: 4
  Publisher: Nature Publishing Group}\BibitemShut {NoStop}%
\bibitem [{\citenamefont {McConville}\ \emph {et~al.}(2020)\citenamefont
  {McConville}, \citenamefont {Lu}, \citenamefont {Wang}, \citenamefont {Tan},
  \citenamefont {Cochard}, \citenamefont {Conroy}, \citenamefont {Moore},
  \citenamefont {Harvey}, \citenamefont {Bangert}, \citenamefont {Chen},
  \citenamefont {Gruverman},\ and\ \citenamefont
  {Gregg}}]{mcconville_ferroelectric_2020}%
  \BibitemOpen
  \bibfield  {author} {\bibinfo {author} {\bibfnamefont {J.~P.~V.}\
  \bibnamefont {McConville}}, \bibinfo {author} {\bibfnamefont
  {H.}~\bibnamefont {Lu}}, \bibinfo {author} {\bibfnamefont {B.}~\bibnamefont
  {Wang}}, \bibinfo {author} {\bibfnamefont {Y.}~\bibnamefont {Tan}}, \bibinfo
  {author} {\bibfnamefont {C.}~\bibnamefont {Cochard}}, \bibinfo {author}
  {\bibfnamefont {M.}~\bibnamefont {Conroy}}, \bibinfo {author} {\bibfnamefont
  {K.}~\bibnamefont {Moore}}, \bibinfo {author} {\bibfnamefont
  {A.}~\bibnamefont {Harvey}}, \bibinfo {author} {\bibfnamefont
  {U.}~\bibnamefont {Bangert}}, \bibinfo {author} {\bibfnamefont {L.-Q.}\
  \bibnamefont {Chen}}, \bibinfo {author} {\bibfnamefont {A.}~\bibnamefont
  {Gruverman}}, \ and\ \bibinfo {author} {\bibfnamefont {J.~M.}\ \bibnamefont
  {Gregg}},\ }\href {\doibase https://doi.org/10.1002/adfm.202000109}
  {\bibfield  {journal} {\bibinfo  {journal} {Advanced Functional Materials}\
  }\textbf {\bibinfo {volume} {30}},\ \bibinfo {pages} {2000109} (\bibinfo
  {year} {2020})},\ \bibinfo {note} {\_eprint:
  https://onlinelibrary.wiley.com/doi/pdf/10.1002/adfm.202000109}\BibitemShut
  {NoStop}%
\end{thebibliography}
%

\end{document}